# Organizational culture and the usage of Industry 4.0 technologies: evidence from Swiss businesses[†]

Simon Alexander Wiese *     Johannes Lehmann **     Michael Beckmann ***

**Abstract**

Using novel establishment-level observational data from Switzerland, we empirically examine whether the usage of key technologies of Industry 4.0 distinguishes across firms with different types of organizational culture. Based on the Technology-Organization-Environment and the Competing Values framework, we hypothesize that the developmental culture has the greatest potential to promote the usage of Industry 4.0 technologies. We also hypothesize that companies with a hierarchical or rational culture are especially likely to make use of automation technologies, such as AI and robotics. By means of descriptive statistics and multiple regression analysis, we find empirical support for our first hypothesis, while we cannot confirm our second hypothesis. Our empirical results provide important implications for managerial decision-makers. Specifically, the link between organizational culture and the implementation of Industry 4.0 technologies is relevant for managers, as this knowledge helps them to cope with digital transformation in turbulent times and keep their businesses competitive.

[†]This research has been funded by the Swiss National Science Foundation as part of the National Research Program NRP77 "Digital Transformation" under grant No. 187462. The authors are solely responsible for the analysis and the interpretation thereof.

* Simon Alexander Wiese, University of Basel, Faculty of Business and Economics, Peter Merian-Weg 6, CH-4002 Basel, Switzerland, phone: +41 61 207 27 54, e-mail: simonalexander.wiese@stud.unibas.ch

*** Johannes Lehmann, University of Basel, Faculty of Business and Economics, Peter Merian-Weg 6, CH-4002 Basel, Switzerland, phone: +41 61 207 58 42, e-mail: j.lehmann@unibas.ch

*** Michael Beckmann, University of Basel, Faculty of Business and Economics, Peter Merian-Weg 6, CH-4002 Basel, Switzerland, Institute for Employment Research (IAB), Nuremberg, Germany, and IZA – Institute of Labor Economics, Bonn, Germany, phone: +41 61 207 32 24, e-mail: michael.beckmann@unibas.ch

## 1. Introduction

Technological progress has been influencing the production of goods and services at all times, sometimes changing them so significantly that, historically speaking, we can now distinguish four industrial revolutions. The current fourth industrial revolution, also referred to as Industry 4.0, is characterized by two main features: the networking of information and communication technologies (ICT) with each other and with production facilities as well as the automation of business and production processes, in which workers are less actively involved in operational tasks than in activities to monitor these production processes. The key technologies of Industry 4.0 include cyber-physical systems, the Internet of Things (IoT), artificial intelligence (AI), big data analytics, cloud computing, robotics, system integration using business software such as enterprise resource planning (ERP) or customer relationship management (CRM), and additive manufacturing (e.g. 3D printing), among others.

While most of the academic literature on Industry 4.0 examines the impact of these technologies on various outcomes, such as employment or productivity, our empirical study focuses on the determinants of the use of Industry 4.0 technologies in organizations. Specifically, we investigate whether firms with different organizational cultures also differ with respect to their use of Industry 4.0 technologies. Our analysis is based on the idea that certain organizational cultures explicitly or at least implicitly promote the use of process innovation such as Industry 4.0 technologies, while other organizational cultures may even hinder process innovation. Furthermore, we explore whether different types of organizational culture promote different types of Industry 4.0 technologies, depending on the degree to which the technology can be considered as disruptive. The objective of our study is therefore to identify process innovation-friendly organizational cultures that make it easier for companies to implement and use Industry 4.0 technologies in order to remain competitive in the future and realize sustaining competitive advantage.

For this purpose, we apply data from the Swiss Employer Survey (SES), which is a new cross-sectional data set containing observational data on a wide range of business topics in Swiss firms, with a focus on Industry 4.0 technologies and management practices. The SES is a primary data set that is based on a random sample provided to us from the Swiss Federal Statistical Office. This sample is representative for Swiss firms with ten or more employees. To increase the sample size, we supplemented the sample from the Swiss Federal Statistical Office with an own sample drawn by means of web scraping. The final data set consists of 498 observations.[1] Methodologically, we rely on descriptive analyses and the estimation of conditional correlations resulting from conventional OLS models, Poisson regression models for count data as well as binary response models. Our general theoretical considerations on organizational culture as a core determinant of Industry 4.0 technology usage in firms build on the Technology-Organization-Environment (TOE) framework developed in DePietro et al. (1990) and described, for example, in Baker (2012), Arnold et al. (2018), and Amin et al. (2024). In terms of a helpful typology of organizational cultures, our theoretical reasoning is based on the well-known Competing Values framework as explained, for example, in Boddy (2019, chapters 2 and 3) as well as McDermott and Stock (1999) and Tortorella et al. (2024).

The empirical literature on the determination (antecedents or enablers) of technological innovations is quite heterogeneous and sometimes does not include organizational culture as a key determinant (e.g. Khin and Kee 2022, Sarbu 2022) or the adoption or usage of Industry 4.0 technologies as an outcome variable (e.g. Saghafian et al. 2021, Nguyen and Aoyama 2014). Moreover, this literature often relies on methods of structural equation modeling (e.g. Martinez-Caro et al. 2020, Nguyen and Aoyama 2014, Amin et al. 2024), qualitative literature reviews (e.g. Saghafian et al. 2021), or qualitative exploratory stu+dy designs based on multiple case studies (e.g. Khin and Kee 2022), thereby missing to account for the endogenous nature of the organizational culture explanatory variable as well as ensuring the generalizability or representativeness of the obtained study results. Our paper contributes to the literature by using observational data with a focus on Industry 4.0 technologies and econometric estimation methods that explicitly address different sources of endogeneity and representativeness, thus allowing a meaningful interpretation of the empirical results.

Our main findings resulting from descriptive statistics and multiple regression analyses reveal a positive association between the developmental culture and the diffusion of Industry 4.0 technologies in Swiss companies, while the remaining types of organizational culture according to the Competing Values framework, i.e., rational culture, hierarchical culture, and group culture. do not turn out to be significantly correlated with the diffusion of Industry 4.0 technologies. This empirical result is in line with our Hypothesis 1, according to which the developmental culture has the greatest potential to be positively associated with the intensity of Industry 4.0 technology usage. In contrast, we find no empirical confirmation for our hypothesis 2, according to which companies with a hierarchical or rational culture are especially likely to make use of automation technologies, such as AI and robotics.

---

[1] For more detailed information on the SES as well as its representativeness and validity in comparison to other observational data sets at the firm level see Lehmann and Beckmann (2024).

Our empirical results provide important implications for managerial decision-makers. Specifically, the link between organizational culture and the implementation of Industry 4.0 technologies is relevant for managers, as this knowledge helps them to cope with digital transformation in turbulent times and keep their businesses competitive.

The remainder of our paper is as follows: In section 2, we briefly describe the TOE framework to explain the theoretical basis of our empirical analysis on the impact of organizational culture on Industry 4.0 technology adoption and usage. Section 2 also introduces the reader to the Competing Values framework representing the theoretical background for distinguishing between the most important types of organizational cultures. Section 3 presents short descriptions of the observational data and the core variables applied in our study. Section 4 contains information on our econometric methodology. In section 5, we present and discuss our empirical results. Finally, section 6 concludes.

## 2. Theoretical background

In this section, we present the theoretical framework for our analysis on the relevance of different forms of organizational culture for firm investments in various key technologies of Industry 4.0. The Technology-Organization-Environment (TOE) framework provides the overarching theoretical framework for this by providing information on the determinants of technological innovation. We consider organizational culture as one of these determinants for the realization of process innovations. Organizational culture itself can be divided into different forms and types. For this purpose, we choose the typology according to the Competing Values framework, which is certainly one of the best-known typologies.

### 2.1 The Technology-Organization-Environment (TOE) framework

According to the TOE framework introduced by in DePietro et al. (1990), an organization's decision to implement or adapt technological innovations depends on three dimensions: the technological, organizational and environmental context of the organization (e.g. Baker 2012, Arnold et al. 2018, Amin et al. 2024). For our analysis, the organizational context is of particular interest as organizational culture is our main explanatory or treatment variable. Figure 1 illustrates the main relationships developed in the TOE framework.

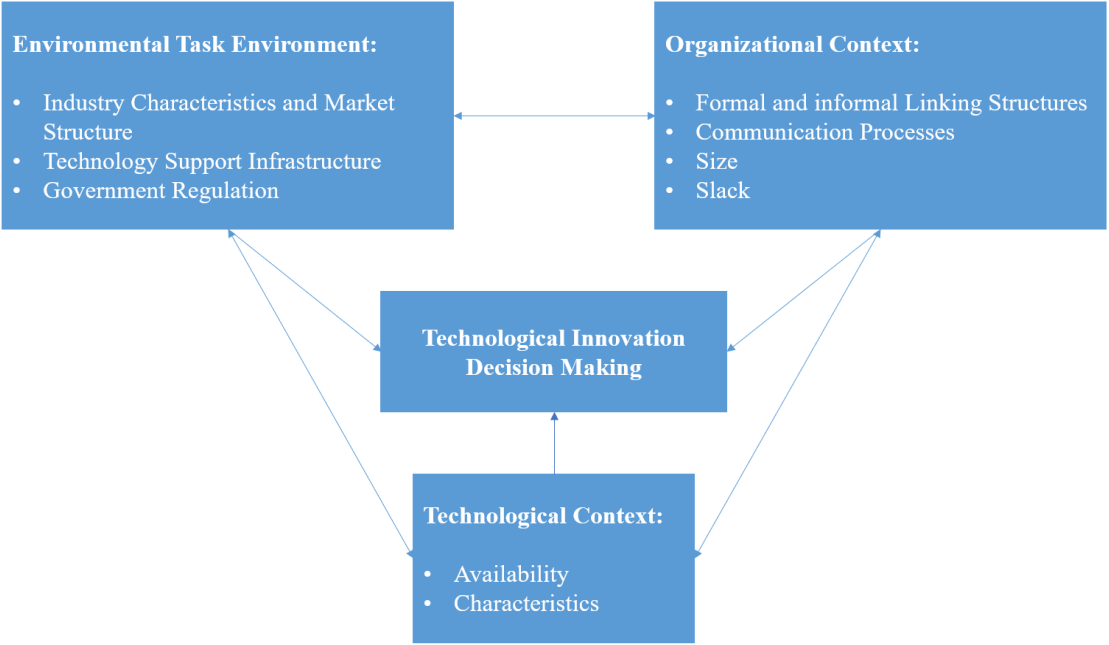

Figure 1: Technology-Organization-Environment framework (adapted from Baker 2012, p. 6, Amin et al. 2024, p. 9)

With regard to the technological context, all technologies that are crucial for the company are important. This includes both technologies that a company already uses and those that it does not yet use, but which could in principle be acquired. While the technologies already available determine the speed and scope of a company's technological change, the technologies not yet in use define the limits of the possibilities and thus indicate the extent to which technologies can support development. Innovations outside the company are described as incremental, synthetic or discontinuous changes. Incremental process innovations offer new features or versions of existing technologies, such as a new version of an ERP system, and therefore pose the least risk and change compared to the other two types of technological change. Synthetic process innovations combine existing ideas and technologies in new ways, as exemplified by the provision of course materials at the university via the Internet. Discontinuous or radical process innovations can be technological innovations that significantly deviate from existing technologies and processes. An example is the switch from mainframe computers to PCs at companies in the 1980s. The availability and characteristics of the technologies are therefore important with regard to the technological context.

With regard to the organizational context, the characteristics and resources of a company are important. This includes the networking structures between employees as well as the communication processes within a company, the size of the company and surplus resources. Mechanisms for establishing connections between individuals and parts of the organization can promote innovation. Examples are informal networking agents, such as boundary spanners or gatekeepers, or cross-functional teams as well as employees with connections to external departments and elements within the value chain. In terms of organizational structure, companies with an organic and decentralized structure are supposed to be advantageous when it comes to the adoption of innovations, not least because they focus on teams that handle the responsibilities of employees in a partially fluid manner and advocate lateral communication alongside traditional, hierarchical communication. In contrast, a classic organizational structure, which is characterized by a focus on formal relationships in reporting, centralized decision-making and clearly defined roles for employees, is better suited to the implementation of innovations. In terms of communication processes, top management can drive innovation through an organizational context that is open to change and promotes innovation that positively impacts the company's core mission and vision. Furthermore, employees should be made aware of the importance of innovation in the context of the overarching strategy and the importance of innovation itself. Top management should also reward innovation formally and informally, build a history of innovation within the organization and assemble a talented leadership team that paints an optimistic picture of the company's future. With regard to the aspects of the size of an organization and the surplus of resources, the literature is inconsistent in terms of its influence on innovation. However, as these are not the main focus of this study, reference will only be made here to the relevant literature. See Baker (2012), March and Simon (1958), Tornatzky et al. (1983), Cyert and March (1963), and Kimberly (1976).

The environmental context deals with the structure of the industry as well as the presence or absence of providers of technological services and the regulations that apply in this environment. In terms of the industry structure, intense competition and, in terms of the industry life cycle, a faster advancing cycle can have a positive influence on innovation adoption. With regard to the availability of technological services, a supply of skilled workers and consultants / technology providers can promote innovation. Legal regulations can have both positive and negative effects on innovation. In the banking sector, for example, rules on data security can mean that new ways of accessing account information cannot be introduced.

## 2.2 The Competing Values framework

The organizational culture can be placed in the organizational context of the TOE framework. Its potential impact on the decision regarding the adoption of new technologies is explained in the course of the work. This framework enables the classification of organizational cultures. The extent to which these cultural types have an influence on the adoption of technologies is examined empirically.

The Competing Values framework distinguishes between four different types of organizational cultures based on two differentiating dimensions (e.g. Denison and Spreitzer 1991, McDermott and Stock 1999, Boddy 2019 (chapters 2 and 3), Cameron and Quinn 2006, Tortorella et al. 2024). The first refers to the tension between flexibility and control, meaning that some companies strive for flexibility to cope with rapid change, while others seek to increase control. The second dimension distinguishes between an internal and an external focus, where the integration to maintain the internal, existing organization has to be distinguished from the focus on competition, adaptation and interaction with the external environment. This is based on the different requirements created by the internal organization and the external environment. Figure 2 illustrates the four types of organizational culture.

|  | **Flexibility** | **Control** |
|---|---|---|
| **Internal Focus** | **Group Culture:**<br><br>• Concern<br>• Commitment<br>• Morale<br>• Discussion<br>• Participation<br>• Openness | **Hierarchical Culture:**<br><br>• Measurement<br>• Documentation<br>• Information Management<br>• Stability<br>• Control<br>• Continuity |
| **External Focus** | **Developmental Culture:**<br><br>• Insight<br>• Innovation<br>• Adaption<br>• External Support<br>• Resource acquisition<br>• Growth | **Rational Culture:**<br><br>• Accomplishment<br>• Productivity<br>• Profit/Impact<br>• Goal Clarification<br>• Direction<br>• Decisiveness |

Figure 2: Competing Values framework (adapted from McDermott and Stock 1999, p. 524, Denison and Spreitzer 1991, pp. 4, 12, Boddy 2019, p. 38, Tortorella et al. 2024, p. 1266)

The group culture is characterized by flexibility and a focus on internal organization and places an emphasis on human relationships. Furthermore, belonging, trust and participation are important values, and leaders are characterized as supportive and considerate. Teamwork supports interaction between members, which emanates from leaders. Connectedness, cohesion and membership are motivating, and the effectiveness of the culture is measured by the development of human potential and the commitment of the group members.

Due to its focus on flexibility and the external environment, a development-oriented culture is geared towards growth, creativity, the acquisition of resources and an adaptation to the environment. Leaders are risk-takers, entrepreneurial and idealistic. They also provide additional resources, visibility, legitimacy and support from external forces. Growth, development of new markets and the acquisition of resources are used to measure effectiveness.

The rational culture is characterized by control and an external focus. Here, value is placed on productivity, performance and the achievement of clearly defined goals. The decision-makers are goal-oriented, functional and direct. They specify structures and drive productivity. In this type of organizational culture, competition and the successful achievement of goals have a motivating effect. Planning, productivity and efficiency are the key effectiveness factors.

Finally, the hierarchical culture is characterized by stability and control as well as an internal focus, which means that internal efficiency, uniformity, coordination and evaluation are important here. Security, rules and regulations have a motivating effect, while managers tend to be conservative and cautious. Effectiveness is assessed on the basis of control, stability and efficiency.

## 2.3 Derivation of the test hypotheses

Our first hypothesis to be tested refers to the relationship between organizational culture and the intensity of Industry 4.0 technology usage. This means that at this stage of the analysis we do not distinguish between certain key technologies of Industry 4.0 but simply consider the firms' amount of technology usage. The theoretical reasoning in section 2.2 suggests that a positive effect on the intensity of the use of Industry 4.0 technologies can basically come from any of the four forms of organizational culture considered according to the Competing Values framework. Above all, however, it is to be expected that the developmental culture promotes the diffusion of Industry 4.0 process innovations. This is because the developmental culture explicitly aims at increasing employee creativity and fostering innovation and new resources.

In contrast, the guiding principles of the other cultural types can be summarized as 'participation fosters commitment' (group culture), 'control fosters efficiency' (hierarchical culture), and 'competition fosters productivity' (rational culture) (Tortorella et al. 2024). This does not exclude positive effects on the implementation of Industry 4.0 technologies, but the idea of innovation is most firmly anchored in the developmental culture. For this reason, we formulate our first hypothesis as:

> *Hypothesis 1: All types of organizational culture within the Competing Values framework have the potential to be positively associated with the intensity of Industry 4.0 technology usage, but the developmental culture has the greatest potential.*

To derive our second hypothesis to be tested, we distinguish between the types of organizational culture focusing on flexibility and autonomy and those types that focus on control and stability. The question here is whether technologies that enable large-scale automation of production and service processes are primarily in demand from companies that place great value on control and stability in their organizational culture, i.e. companies with a hierarchical or rational culture. We argue that organizational cultures emphasizing values such as goals and productivity or control and efficiency are supportive when it comes to the acquisition of automation technologies such as AI or robotics, while firms with a developmental or a group culture are supposed to focus on improvements in production or service delivery rather than replacing workers. Hence, our second hypothesis to be tested is:

> *Hypothesis 2: Companies with a hierarchical or rational culture are more likely to make use of automation technologies, such as AI and robotics, than companies with a developmental or group culture.*

In section 5, we present and discuss some empirical tests on *Hypotheses 1* and *2* resulting from our econometric analysis.

## 3. Data, variables, and descriptive statistics

In this section, we introduce the reader to the cross-sectional survey data on which our research project is based. We then present our dependent, explanatory and control variables and provide some descriptive statistics to give the reader a first impression of our empirical analysis.

## 3.1 The Swiss Employer Survey (SES)

Our empirical study is based on the Swiss Employer Survey (SES), which is a primary data set collected from establishments in the Swiss economy.[2] The SES is a cross-sectional data set providing observational data for the years 2020 or 2022, respectively. The addresses of the establishments were made available to us by the Swiss Federal Statistical Office in the form of a sample that is representative of Swiss establishments with ten or more employees.[3] In total, we were able to contact 10,000 establishments in this way and asked for their support of our research project. To ensure sufficient coverage of larger establishments with 250 or more employees, care was taken that this group was disproportionately represented in the drawing of the initial sample.

The SES covers a wide range of business information on workforce structures, organizational cultures and corporate strategies, decision-rights assignment, performance measurement, remuneration policies, the establishments' financial situation as well as their business environment including their market situation and regulation issues. Other topics refer to staff recruiting, working time regimes and further training. The focus of the SES, however, is on where Swiss establishments stand in terms of digital transformation. This includes information on the adoption and usage of a wide range of Industry 4.0 technologies, such as cyber-physical systems, the Internet of Things (IoT), artificial intelligence (AI), big data analytics, cloud computing, robotics, horizontal and vertical system integration using business software such as enterprise resource planning (ERP) or customer relationship management (CRM), and additive manufacturing (e.g. 3D printing).

To increase the number of observations resulting from the responses to our initial survey (322 establishments), we supplemented the initial sample provided by the Swiss Federal Statistical Office with a data set drawn via web scraping.[4] In this way, we were able to increase the number of observations by 176 establishments. In the end, our final data set consists of 498 firms. Lehmann and Beckmann (2024), who also make use of the SES observational data, demonstrate that despite the relatively low response rate (caused by launching the data collection during the height of the Covid-19 pandemic) and the merge of two establishment-level data sets, there is a great deal of comparability between the initial and the web-scrapped sample and between the merged SES data and other firm-level survey data sets from Switzerland and abroad. Overall, therefore, there is no indication that would justify questioning the validity or representativeness of our SES observational data.

---

[2] The introduction to the data set used for our empirical analysis draws heavily on Lehmann and Beckmann (2024), where the reader can find a more detailed description of the data.
[3] The initial sample does not include information from establishments of the public administration, farming, and mining sectors.
[4] The first step to construct the web scraped sample was to manually compile a list of employer and industry affiliations. In a second step, the contact details of the member establishments listed on the organization's website were extracted either manually (for smaller organizations) or automatically using a Python script (for larger organizations). We contacted the establishments of the generated sample by e-mail. To keep the survey population comparable to the baseline sample, we excluded establishments with less than ten employees and establishments that already participated in our baseline survey.

**3.2 Industry 4.0 technology variables**

Our SES data contains information on the adoption and usage of a wide range of Industry 4.0 technologies in Swiss businesses. In total, the SES provides information on 14 types of Industry 4.0 technologies. Figure 3 presents a summarizing description of all 14 binary technology variables $I4.0^1, I4.0^2, ..., I4.0^{14}$.

The use of groupware (around 80%) and cloud computing (around 63%) is relatively widespread in Swiss companies. This applies to both the manufacturing sector and the service sector. Differences regarding the incidence of Industry 4.0 technologies across sectors, however, can be observed for the business software solutions allowing the horizontal and vertical system integration. What is particularly striking here is that the incidence of ERP is somewhat greater in the manufacturing sector (68%) than in the service sector (57%), while the opposite is true for DMS, CRM and MIS. It is also noticeable that MIS is less widespread in Swiss companies than the other business software solutions (18% manufacturing sector, 26% service sector). Even though public media reporting today can easily give the impression that artificial intelligence (AI) and big data analytics are already playing a dominant role, at least in larger companies, the actual diffusion rate is only 36% in the manufacturing sector and 23% in the service sector. Other Industry 4.0 technologies are also used less in Swiss companies than one might have expected. This applies to the Internet of Things (19% manufacturing sector, 14% service sector), virtual boardrooms (17% manufacturing sector, 16% service sector), robotics (21% manufacturing sector, 5.5% service sector), additive manufacturing / 3D print (15.5% manufacturing sector, 3% service sector), virtual / augmented reality (5% manufacturing sector, 3.5% service sector) and blockchain technology (0% manufacturing sector, 1.5% service sector). At first glance, it seems somewhat unusual that cyber-physical systems are more widespread in the service sector than in the manufacturing sector, while the reverse is true for the IoT. This can at least partially be explained by the fact that the two terms are sometimes regarded as synonyms. Overall, therefore, figure 3 also demonstrates that Industry 4.0 technologies are not just a phenomenon of the manufacturing sector. With a few exceptions (additive manufacturing, robotics), the key technologies of Industry 4.0 are also prevalent in the Swiss service sector.

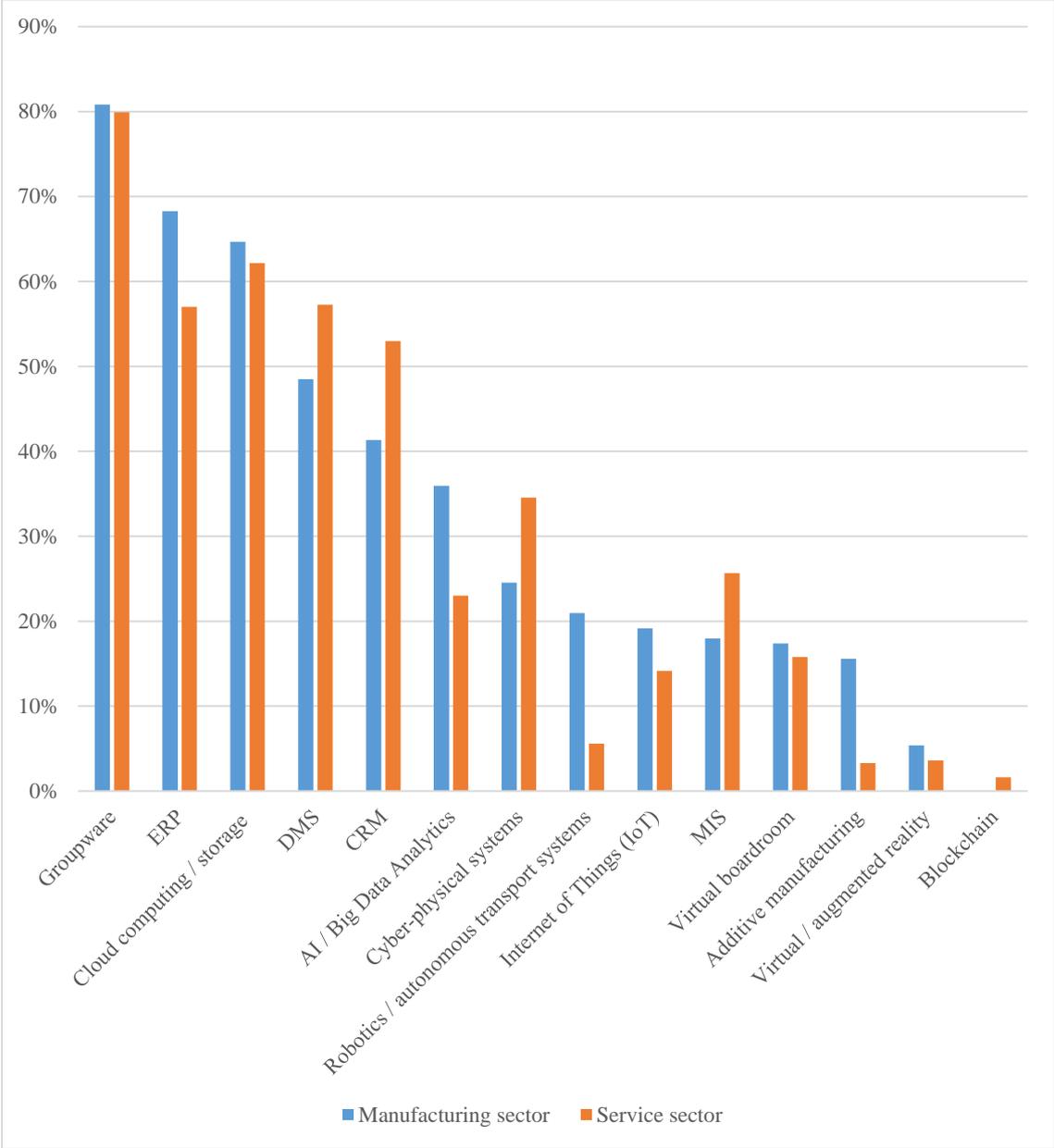

Figure 3: Incidence of the key technologies of Industry 4.0 in Swiss establishments. Source: Swiss Employer Survey (SES).

We apply each of the binary technology variables as a dependent variable. In addition, we construct a composite variable $I4.0^{INT1}$ that provides information on the intensity of Industry 4.0 technologies usage, thus ensuring a comprehensive view of the state of digital transformation in Swiss companies. For the construction of this composite index variable, we follow empirical studies, such as Bresnahan et al. (2002), Bloom et al. (2011), Gerten et al. (2019, 2022), as well as Beckmann and Kräkel (2022), and apply a double-standardization approach, so that the resulting technology intensity variable $I4.0^{INT1}$ can thus be written as

$$I4.0^{INT1} = STD\{STD(I4.0^1) + STD(I4.0^2) + \cdots + STD(I4.0^{14})\}.$$

By construction, $I4.0^{INT}$ has zero mean and unit variance.

Another outcome variable measuring the intensity of Industry 4.0 technology usage is simply the sum of all individual technologies that are currently in use in Swiss companies, i.e.,

$$I4.0^{INT2} = I4.0^1 + I4.0^2 + \cdots + I4.0^{14}.$$

Obviously, $I4.0^{INT2}$ is a count variable ranging between 0 and 14, i.e., $I4.0^{INT2} \in [0,14]$. Figure 4 shows the distribution of $I4.0^{INT2}$.

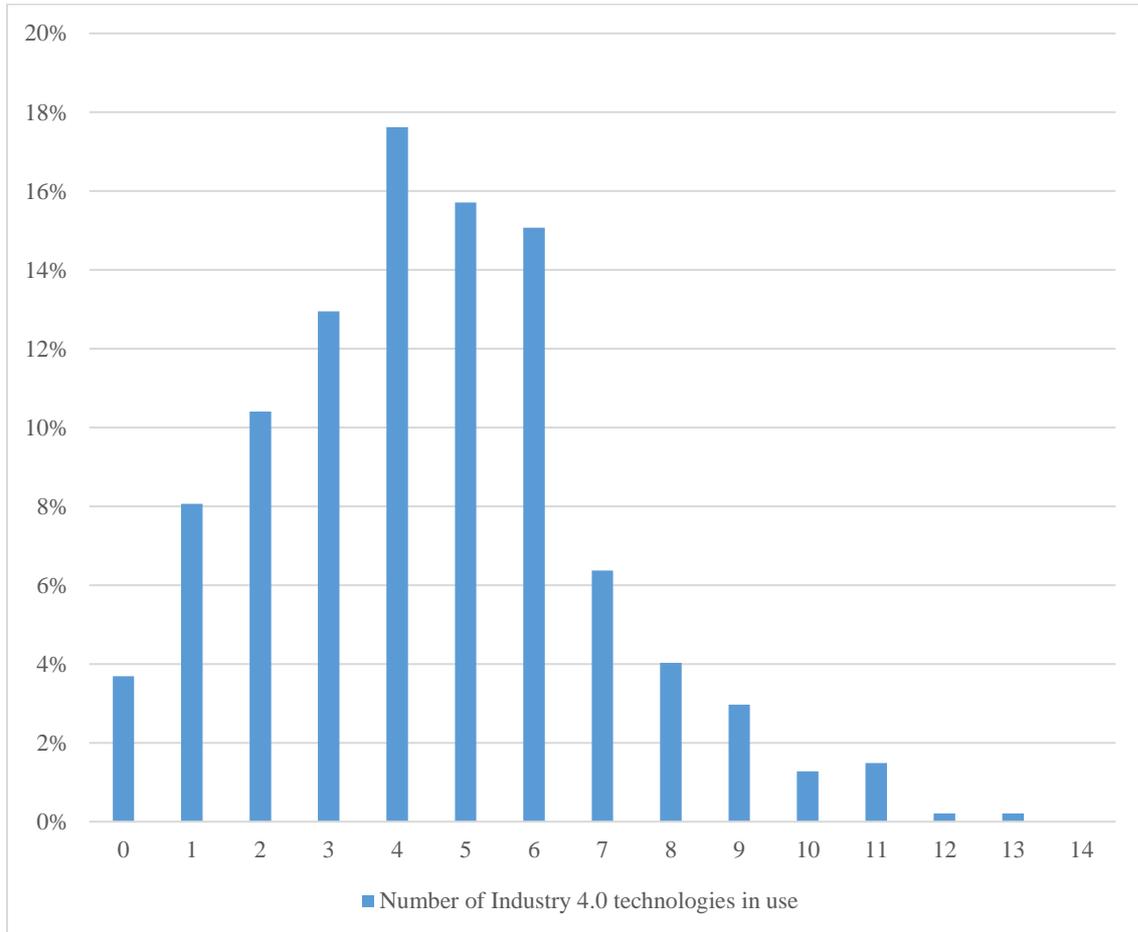

Figure 4: Number of the key technologies of Industry 4.0 used in Swiss establishments. Source: Swiss Employer Survey (SES).

Figure 4 shows that very few companies are currently not using any of the key technologies of Industry 4.0 (less than 4%). The modal value is four key technologies in use (just under 18%). However, there are also some companies that use seven or more Industry 4.0 technologies.

### 3.3 Organizational culture variables

We measure the types of organizational culture based on the Competing Values framework by making use of firm-specific information on the importance of certain dimensions of organizational culture. These dimensions are

a) Focus on employee performance / outcomes ($PERF$): the extent to which managers focus on results rather than on how these results were achieved,

b) Focus on employees ($EMP$): the extent to which management decisions take into account the impact on the organization's employees,

c) Focus on team work ($TEAM$): the extent to which work is organized to be completed by teams rather than by individuals,

d) Focus on competition ($COMP$): the extent of aggressiveness and competitive orientation of employees instead of cooperative behavior,

e) Focus on stability ($STAB$): the extent to which decisions and actions of the organization adhere to the status quo,

f) Focus on innovation and risk taking ($INNO$): extent to which employees are encouraged to innovate and take risks,

g) Focus on precision ($PREC$): the extent to which employees are expected to work accurately, analyze, and pay attention to detail.

The possible firm responses are measured at an ordinal scale ranging between 1 (unimportant) to 5 (important). Figure 5 reports the descriptive statistics of the considered types or organizational culture.

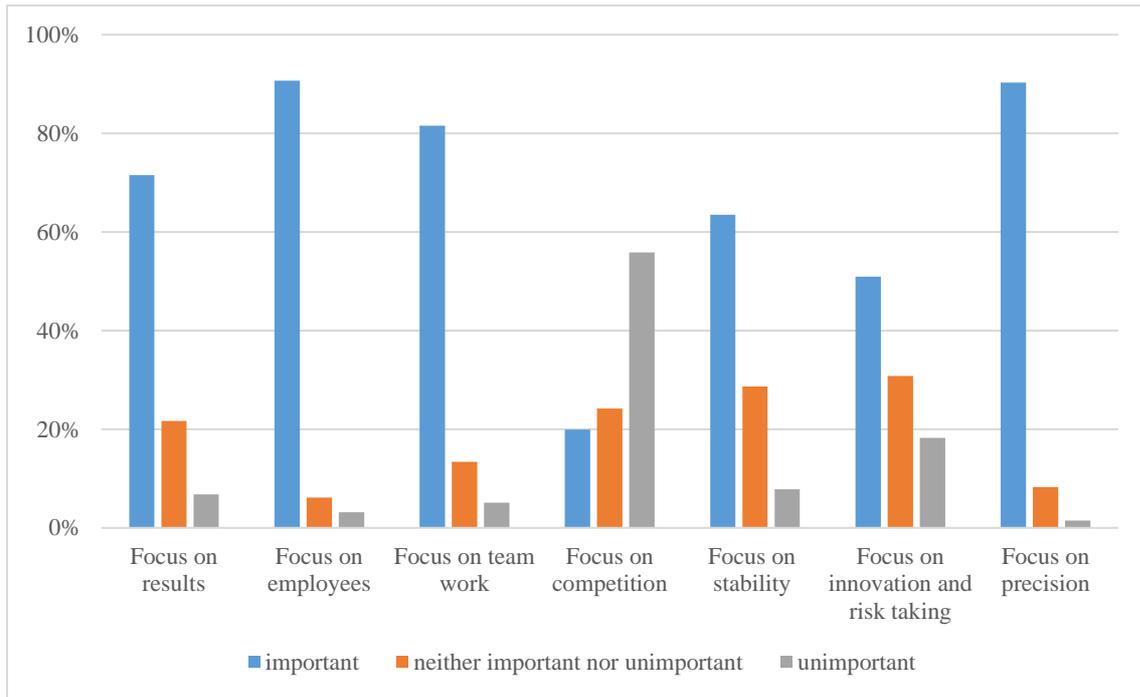

Figure 5: Importance of the considered types of organizational culture in Swiss establishments. Source: Swiss Employer Survey (SES).

Using the double-standardization approach, we define our variables measuring the different types of organizational culture as follows:

- Group culture (collaborate culture):

$$OC^{group} = STD\{STD(TEAM) + STD(EMP)\}.$$

- Rational culture (competitive culture):

$$OC^{rat} = STD\{STD(PERF) + STD(COMP)\}.$$

- Hierarchical culture (controlling culture):

$$OC^{hier} = STD\{STD(STAB) + STD(PREC)\}.$$

- Developmental culture (creative culture):

$$OC^{dev} = STD\{INNO\}.$$

All $OC$ variables have zero mean and unit variance by construction. We utilize the standardized $OC$ variables to construct binary treatment variables that separate strong from weak organizational cultures. More precisely, we use the zero mean of the respective $OC$ distribution and define our binary explanatory variables as

$$OCD^l = \begin{cases} 1 & \text{if } OC^l > 0 \\ 0 & \text{if } OC^l \leq 0 \end{cases},$$

where $l \in \{group, rat, hier, dev\}$.

### 3.4 Control variables

To estimate informative and meaningful effects of organizational culture on Industry 4.0 technology usage, our econometric modeling must include carefully selected, theory-driven control variables. An econometric model without appropriate control variables is unlikely to produce valid parameter estimate but is at risk of suffering from a severe omitted variables bias. The TOE framework is not only the theoretical basis for explaining the relationship between organizational culture and Industry 4.0 technology usage. Due to its broad orientation, the TOE framework does also provide the theoretical background for the choice of the control variables.

Managerial decisions on the implementation of process innovations naturally depend on the existing technological equipment or status of the companies pointing at the relevance of the technological context of the TOE framework for the determination of Industry 4.0 technology usage. Our first technology-related control variable therefore is a variable providing information about the technological state of the company. This variable is the standardized version of a self-reported variable at an ordinal scale ranging between 1 (technologies in use are out of date) to 5 (technologies in use are state-of-the-art). Furthermore, in the models, where we estimate the determinants of the individual Industry 4.0 technology usage $I4.0_i^j$, we control the effects of the remaining Industry 4.0 technologies $I4.0_i^{k \neq j}$.

We map the organizational context of the TOE framework by an industry dummy separating the manufacturing from the service sector and by two dummy variables indicating a company's make-or-buy strategies (internal and external expansion strategies, business unit sales and outsourcing decisions) in the past five years. Furthermore, we control for company size by three dummy variables (small, medium-sized, large) and for the skill structure (share of high- and medium-skilled employees relative to total workforce) within companies. Finally, we control for organizational leadership by two dummy variables indicating whether the company is managed by the founders or owners of the organization or by paid managers.

To capture the environmental domain of the TOE framework, we control for the number of competitors as a measure of competitive pressure and for the seven Swiss greater regions (e.g., Northwestern Switzerland, Zurich, Central Switzerland, Ticino) to reflect different local demand conditions. The regulatory component of the environmental context is measured by the legal form of a company (private vs. capital company), the existence of an organizational unit for employee representation (similar to a works council or firm-level union), and a dummy variable indicating whether or not a company is legally independent or part of a larger organization.

## 4. Empirical methodology

The starting point of our empirical analysis is a multiple regression model of the form

$$I4.0_i^{INT} = \sum_l \alpha^l \, OCD_i^l + X_i \beta_1 + u_i \,, \tag{1}$$

where $I4.0_i^{INT} \in \{I4.0_i^{INT1}, I4.0_i^{INT2}\}$ and $i$ indexes firm $i$. Both dependent variables are defined in section 3.2. If $I4.0_i^{INT} = I4.0_i^{INT1}$, equation (1) represents a conventional OLS model. If $I4.0_i^{INT} = I4.0_i^{INT2}$, equation (1) is specified as a count data model. In the latter case, we estimate equation (1) by making use of a Poisson model.

In equation (1), $OCD_i^l$ denotes the four binary variables defined in section 3.3, representing the types of organizational culture according to the competing values model, i.e., $OCD^{group}$, $OCD^{rat}$, $OCD^{hier}$, and $OCD^{dev}$. Furthermore, $X$ is the matrix of control variables discussed in section 3.4, while the $\alpha^l$-scalars and the vector $\beta_1$ represent the parameters to be estimated, where $\alpha^l$ are the parameters of interest. Finally, $u_i$ is a stochastic error term with zero mean and finite variance. Equation (1) allows us to test *Hypothesis 1*, according to which $\alpha^{dev} > 0$ and $\alpha^{dev} \geq \alpha^{l \neq dev}$.

To empirically test *Hypothesis 2*, according to which we primarily expect the developmental organizational culture to promote the firms' usage of any Industry 4.0 technology, while we expect the rational organizational culture to promote only the usage of disruptive technologies such as AI and robotics, we specify the following binary response model:

$$I4.0_i^j = \sum_l \alpha^l \, OCD_i^l + X_i \beta_1 + I4.0_i^{k \neq j} \beta_2 + u_i \,. \tag{2}$$

Here, $I4.0_i^j$ represents the dependent variable on the use of a certain key technology $j$ of Industry 4.0 in firm $i$. Furthermore, $I4.0_i^{k \neq j}$ denotes the matrix of all remaining key technologies of Industry 4.0 except $I4.0_i^j$. The inclusion of the regressor matrix $I4.0_i^{k \neq j}$ allows for the possibility of complementary Industry 4.0 technologies, meaning that many firms are unlikely to adopt or use only on key technology of Industry 4.0. Instead it is more realistic to assume that firms implement or use two or more Industry 4.0 technologies because Industry 4.0 is intended to be used as a complete production system in which various information and production technologies must be interconnected in order to be able to unfold the full effect (Cho et al. 2023).

The $\alpha^l$-parameters in equations (1) and (2) can be interpreted in terms of causal inference only when the variables on organizational culture, i.e., the $OCD_i^l$-variables, satisfy the assumption of conditional independence, often also referred to as unconfoundedness. This excludes any endogeneity issues, such as unobserved confounding, reverse causation or simultaneity, selectivity, and measurement error. As a result, conditional independence is a very critical assumption. This is because at the present stage of data availability, we can only apply observational data at the cross-sectional level. However, in order to convincingly estimate causal effects of organizational culture on Industry 4.0 technology usage without applying an instrumental variable estimation strategy[5], the availability of a comprehensive panel data set including pre-treatment control variables and post-treatment outcome variables would be required. In this case, we could apply a selection-on-observables identification strategy by estimating average treatment effects. In the absence of panel data, however, it appears unlikely that our parameter estimates for the variables representing the four types or organizational culture according to the Competing Values framework can be interpreted in terms of causal inference. As a consequence, our estimation results represent conditional correlations rather than causal effects. Nevertheless, we make some first steps to address different aspects of endogeneity.

This is because it appears reasonable to assume that, despite the cross-sectional nature of our observational data, our estimates are unlikely to suffer from reverse causation or simultaneity bias as one important source of endogeneity. The reason for this assumption is that, most likely, the organizational culture in a company has already existed (significantly) longer than the re-spective Industry 4.0 technologies in use. An organizational culture cannot be developed overnight. It often takes many years before the values and norms of the company founders can be established in the form of an organizational culture in a company. The development of an organizational culture is therefore dynamic and consequently path-dependent.

---

[5] Instrumental variables (IV) estimation would be a suitable selection-on-unobservables method to meet the conditional independence assumption even for analyses with cross-sectional data. The major challenge for IV estimates typically lies in finding a valid IV. In our case, it is unrealistic to find IVs that are both relevant and exogenous, especially given the four variables on organizational culture to be instrumented.

This does not apply in this form to process innovations and technical change in general. For example, in contrast to organizational innovations, technological innovations are usually not regarded as a quasi-fixed input factor in a production function. Technological innovations can be subject to continuous adjustments. In this respect, it is very likely that a company has only recently adopted one or more key technologies of Industry 4.0, while the organizational culture supporting these technologies has already been established in the company for many years.

The idea to mitigate the consequences of endogeneity problems for our empirical analysis has an equally vital relevance when it comes to the selection of our control variables. Here, we also apply the principle of chronological order in order to avoid a bad controls problem. For example, the company representatives in our questionnaire are asked to provide information on corporate strategies dating back up to five years. It can therefore also be expected for this control variable that the corporate strategy took place before the implementation of any Industry 4.0 technologies. Finally, it can also be expected for the remaining control variables that the respective interventions took place earlier than the implementation of Industry 4.0 technologies. Examples include the variables relating to the choice of legal form of an organization and the choice of industry or location.

Overall, therefore, the likely temporal sequence in the implementation of the corporate decisions underlying our dependent and explanatory variables helps us to rule out reverse causation as a cause of endogeneity with all confidence. Furthermore, with the choice of our control variables, we aim at avoiding a bad controls problem. Despite these efforts, we still cannot estimate credible cause-and-effect relationships, but our conditional correlations should be relieved of an important part of the endogeneity problem.

## 5. Empirical results

In this section, we empirically test the association between organizational culture and the use of the key technologies of Industry 4.0 in the Swiss economy, thereby proceeding in two steps. First, we provide some descriptive evidence on this relationship. In a second step, we present the estimation results of our regression analysis introduced in section 4.

## 5.1 Organizational culture and Industry 4.0 technology diffusion: descriptive evidence

Figure 6 indicates descriptive evidence regarding the relationship between the four types of organizational culture according to the Competing Values framework and the diffusion of Industry 4.0 technology usage in Swiss businesses, where technology diffusion is measured by the double-standardized dependent variable $I4.0^{INT1}$.

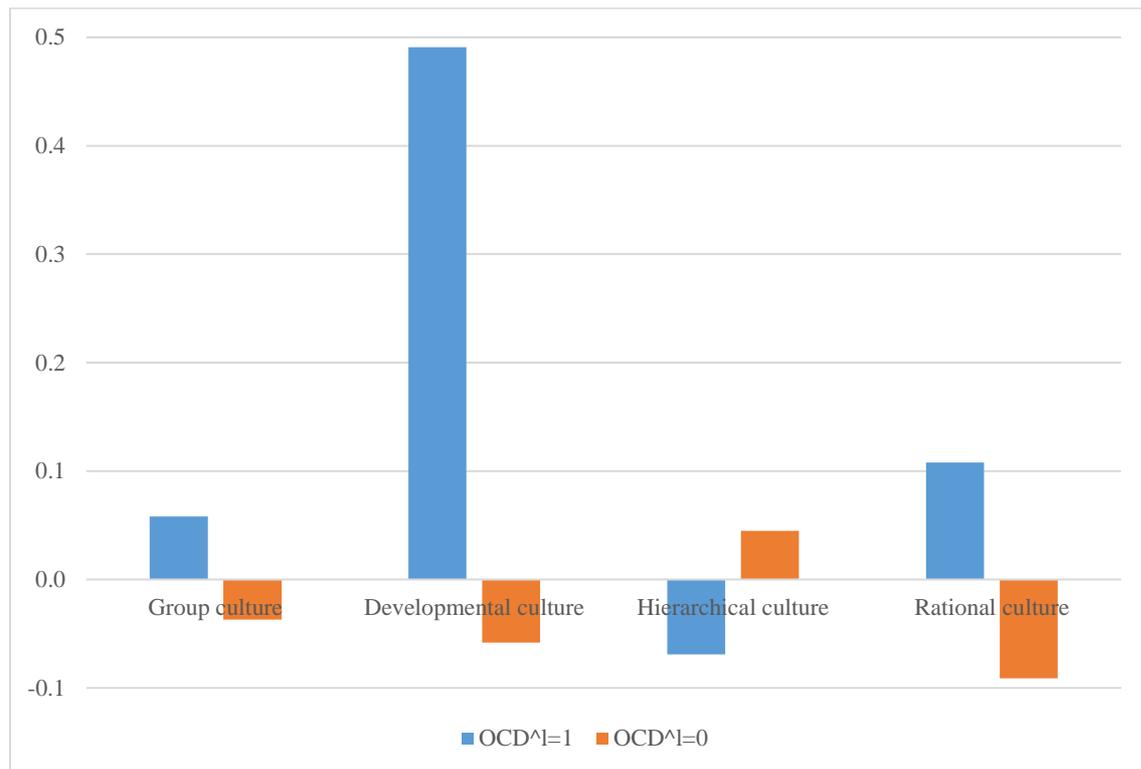

Figure 6: Association between organizational culture and Industry 4.0 technology diffusion measured by $I4.0_i^{INT1}$ in Swiss establishments. Source: Swiss Employer Survey (SES).

It becomes visible that the double-standardized $I4.0^{INT1}$-values differ substantially depending on whether a culture type can be described as strong ($OCD^l = 1$) or weak ($OCD^l = 0$). For three out of the four culture types (group culture, developmental culture, rational culture), the $I4.0^{INT1}$-values are positive (negative), when the respective culture type is considered as strong (weak), while it is just the opposite for the remaining type of the hierarchical culture. This suggests a positive relationship between organizational culture and Industry 4.0 technology diffusion for the group, the developmental and the rational culture types.

Figure 7 expresses the same relationship as figure 6, with the difference that $I4.0^{INT1}$ is replaced here by $I4.0^{INT2}$.

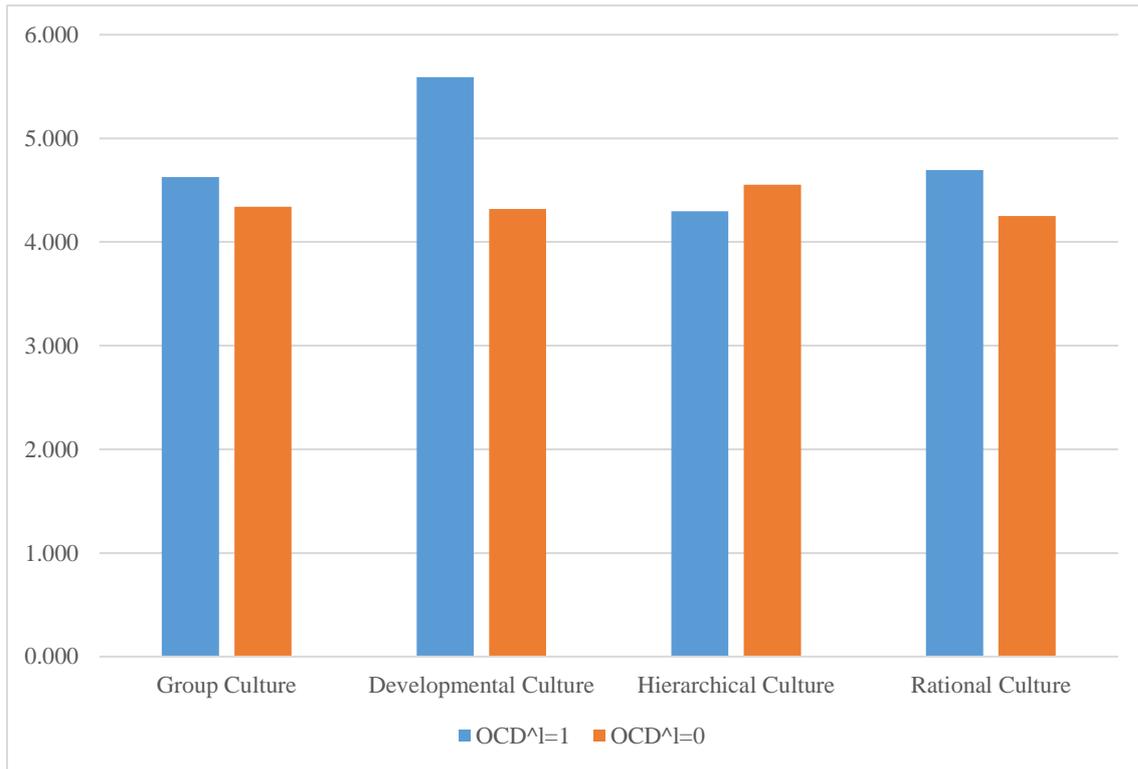

Figure 7: Association between organizational culture and Industry 4.0 technology diffusion measured by $I4.0_i^{INT2}$ in Swiss establishments. Source: Swiss Employer Survey (SES).

The figure illustrates larger differences across strong and weak organizational cultures only for the developmental culture. Here, the mean number of Industry 4.0 technologies is around 5.5 for firms with a strong developmental culture, whereas the mean is around 4.2 for firms with a weak culture. These descriptive finding may be viewed as an indication for the validity of *Hypothesis 1,* according to which the developmental culture type is most likely to be positively associated with the diffusion of Industry 4.0 technologies, but the difference between strong and weak rational and hierarchical cultures is relatively small, so it appears at least questionable whether both culture types can be found to be positively associated with the diffusion of automation technologies, which would be consistent with *Hypothesis 2*. However, in order to make more reliable claims in terms of informative and meaningful associations between organizational culture and Industry 4.0 technology diffusion, this descriptive evidence is not sufficient. For this reason, we go one step further with our empirical analysis and present the results of our regression analysis in the next section.

**5.2 Organizational culture and Industry 4.0 technology diffusion: regression results**

The regression results for equation (1) are displayed in Table 1.

Table 1: Organizational culture and the intensity of Industry 4.0 technologies

| Estimation model | OLS | Poisson regression |
|---|---|---|
| Dependent variable | Intensity of Industry 4.0 technology usage | |
| | $I4.0_i^{INT1}$ | $I4.0_i^{INT2}$ |
| Explanatory variables | (1) | (2) |
| Group culture ($OCD^{group}$) | 0.036 | 0.025 |
| | (0.073) | (0.045) |
| Rational culture ($OCD^{rat}$) | 0.097 | 0.048 |
| | (0.079) | (0.045) |
| Hierarchical culture ($OCD^{hier}$) | −0.135* | −0.064 |
| | (0.076) | (0.046) |
| Developmental culture ($OCD^{dev}$) | 0.308** | 0.134** |
| | (0.134) | (0.065) |
| $N$ | 498 | 498 |

*Note*: Swiss Employer Survey (SES). The values in parentheses represent robust standard errors. $N$ is the number of observations. All specifications additionally contain an identical set of covariates described in section 3.4. * $p < 0.10$; ** $p < 0.05$; *** $p < 0.01$.

The regression results show that of the four cultural types according to the Competing Values framework, only the developmental culture is positively correlated with the intensity of use of Industry 4.0 technologies. This applies to both specifications the OLS model with $I4.0_i^{INT1}$ as dependent variable and the Poisson model with $I4.0_i^{INT2}$ as dependent variable.[6] By contrast, the estimated coefficients for the group and the rational culture prove to be positive, but are not statistically significant. Interestingly, in the OLS specification with the double-standardized $I4.0_i^{INT1}$-variable as dependent variable, the hierarchical culture turns out to be negatively related to the intensity of technology usage, indicating that an organizational culture with a strong orientation on values, such as documentation, stability and control, may tend to be too conservative when it comes to implementing process innovations such as Industry 4.0 technologies. In the Poisson model with $I4.0_i^{INT2}$ as the dependent variable, we also obtain a negative coefficient for $OCD^{hier}$, but this is not statistically significant ($p = 0.163$). This ambiguous estimation result for the culture variable $OCD^{hier}$ is consistent with the corresponding descriptive statistics in section 5.1. There is indeed some indication of the existence of a negative correlation between the hierarchical culture and the intensity of Industry 4.0 technology usage if we look at the descriptive results in figures 6 and 7. Overall, however, we can state that our empirical evidence is consistent with *Hypothesis 1*.

---

[6] The values of the Akaike information criterion (AIC) are practically identical for the Poisson model ($AIC = 2,106$) and the negative binomial model ($AIC = 2,108$), which speaks more in favor of using the Poisson model. This recommendation is confirmed by the result of a likelihood ratio test, which cannot reject the null hypothesis that the data-generating process follows a Poisson distribution ($p = 0.9384$).

Table 2: Organizational culture and the use of individual Industry 4.0 technologies

| Estimation model | Probit ML | | | |
|---|---|---|---|---|
| Explanatory variables | Group culture ($OCD^{group}$) | Hierarchical culture ($OCD^{hier}$) | Rational culture ($OCD^{rat}$) | Developmental culture ($OCD^{dev}$) |
| Dependent variable | (1) | (2) | (3) | (4) |
| AI / Big Data | −0.045 | −0.058 | 0.197 | 0.503** |
|  | (0.139) | (0.140) | (0.136) | (0.211) |
| ERP | −0.136 | −0.346** | −0.296** | 0.069 |
|  | (0.139) | (0.141) | (0.140) | (0.220) |
| DMS | 0.170 | −0.338** | −0.049 | −0.200 |
|  | (0.137) | (0.140) | (0.134) | (0.203) |
| MIS | 0.124 | 0.146 | 0.114 | 0.209 |
|  | (0.141) | (0.149) | (0.147) | (0.198) |
| CRM | 0.182 | −0.047 | 0.151 | 0.287 |
|  | (0.131) | (0.130) | (0.127) | (0.206) |
| Groupware | 0.103 | −0.169 | 0.058 | 0.008 |
|  | (0.163) | (0.166) | (0.166) | (0.233) |
| Virtual boardroom | −0.049 | −0.061 | 0.224 | 0.226 |
|  | (0.161) | (0.169) | (0.168) | (0.227) |
| Cloud comp. / storage | 0.268** | −0.137 | 0.382*** | 0.100 |
|  | (0.130) | (0.130) | (0.125) | (0.214) |
| Cyber-physical system | 0.026 | 0.128 | −0.089 | 0.250 |
|  | (0.142) | (0.138) | (0.135) | (0.208) |
| Internet of Things | 0.070 | 0.017 | −0.094 | 0.482** |
|  | (0.165) | (0.166) | (0.166) | (0.237) |
| Robotics | −0.180 | 0.018 | −0.010 | −0.047 |
|  | (0.205) | (0.221) | (0.213) | (0.287) |
| Additive manufacturing | −0.348 | 0.010 | 0.213 | 0.144 |
|  | (0.229) | (0.218) | (0.220) | (0.315) |
| Virt. / augmented reality | 0.155 | −0.579 | −0.132 | 0.436 |
|  | (0.323) | (0.382) | (0.325) | (0.433) |
| Blockchain | −1.186 | −5.657 | 1.025 | 1.202 |
|  | (2.054) | (7.130) | (2.664) | (4.897) |
| $N$ | 498 | 498 | 498 | 498 |

*Note*: Swiss Employer Survey (SES). The values in parentheses represent robust standard errors. $N$ is the number of observations. All specifications additionally contain an identical set of covariates described in section 3.4. * $p < 0.10$; ** $p < 0.05$; *** $p < 0.01$.

Table 2 displays the regression results for equation (2) that aims at testing *Hypothesis 2*. The estimation results in table 2 do not reveal any new findings overall, except that the positive correlation between the developmental culture and the use of Industry 4.0 technologies is primarily driven by AI / big data analytics and the IoT. Otherwise, it is interesting to note the positive influence on cloud computing by opposing cultural types such as group culture and rational culture, as well as the negative effect of hierarchical culture on ERP and DMS. The latter result once again indicates that a conservative organizational culture is an obstacle to process innovation, which in the age of digital transformation can result in competitive disadvantages for the companies concerned that should not be underestimated.

## 6. Conclusion

In this paper, we empirically examine the question of whether companies find it easier to implement and utilize key technologies of Industry 4.0 if they follow certain types of organizational cultures in their companies. With regard to the types of organizational culture we refer to the typology of the Competing Values framework, which distinguishes between four distinct types of organizational culture, namely the rational culture, the hierarchical culture, the group culture, and the developmental culture. From the theoretical viewpoint, our research question is based on the Technology-Organization-Environment (TOE) framework, according to which technology adoption and usage in companies is determined by their technological, organizational, and environmental surroundings.

To shed light on the empirical relationship between organizational culture and the usage of Industry 4.0 key technologies, we make use of novel observational data at the establishment level. Due to the cross-sectional nature of our survey data set, we do not aim at estimating cause-and-effect relationships. Instead we aim at estimating informative and meaningful conditional correlations that allow us at least to mitigate the issues of reverse causation and bad controls. To mitigate reverse causation, we exploit the fact that the organizational culture in companies was most likely implemented (long) before the adoption of Industry 4.0 technologies. In a similar vein, we argue to avoid the bad controls problem by relying on control variables that are unlikely to be the outcome of a particular firm intervention or that occurred at the same time or after the events of culture implementation or technology adoption. Insofar, we believe that our estimation results obtained from OLS, Poisson and binary response models are more reliable than conventional conditional correlations.

Our estimation results show that the diffusion of Industry 4.0 technologies is higher, when firms follow a developmental culture that values innovative behavior and risk-taking. We do not find evidence consistent with the hypothesis that firms with a hierarchical or rational culture are more likely to invest in automation technologies such as AI or robotics. The values of a hierarchical or rational organizational culture would have fitted in with the use of automation technologies, but we cannot substantiate this consideration.

Our empirical results provide important management implications. Specifically, the link between organizational culture and the implementation of Industry 4.0 technologies is relevant for managers, as this knowledge helps them to cope with digital transformation und turbulent times and keep their businesses competitive. Moreover, we come to the conclusion that in Swiss companies, the implementation of key Industry 4.0 technologies appears to focus less on automation at the expense of labor than is often feared.